%% file: CompileTimeSymbolicAD.tex
\pdfoutput=1

\documentclass[a4paper,12pt]{article}
\usepackage[T1]{fontenc}
\usepackage{lmodern}
\usepackage{lipsum}

\usepackage[ruled]{algorithm2e}

\usepackage[usenames]{color}
\usepackage{amsmath}
\usepackage{amssymb,amsthm}
\usepackage{graphicx}
\usepackage{caption}
\usepackage{authblk}
\usepackage[top=2cm, bottom=2cm, left=2cm, right=2cm]{geometry}
\usepackage{fancyhdr}
\usepackage{url}
\newtheorem{remark}{Remark}

\usepackage{listings}
\SetAlFnt{\small}
\SetAlCapFnt{\small}
\SetAlCapNameFnt{\small}
\SetAlCapHSkip{0pt}
\IncMargin{-\parindent}

\newcommand{\Squeeze}{\mathcal{S}}

\newcommand{\pp}{++}
\newcommand{\Cpp}{C\pp}
\newcommand{\FADBADpp}{FADBAD\pp}

\newcommand{\gnunew}{\textbf{GNU}}

\newcommand{\sun}{\textbf{Sun}}

\newcommand{\approachint}{\textbf{Integer}}
\newcommand{\approachreal}{\textbf{Real}}

\newcommand{\typeof}{{\lstinline+decltype+}}
\newcommand{\auto}{{\lstinline+auto+}}

\newcommand{\midrule}{\noalign{\global\arrayrulewidth1pt}\hline
                      \noalign{\global\arrayrulewidth0.5pt}\rule{0pt}{12pt}}

\begin{document}



\title{\sf Compile-Time Symbolic Differentiation Using
       C\raisebox{0.5ex}{\large++} Expression Templates}

\author[1]{Drosos Kourounis}
\author[2]{Leonidas Gergidis}
\author[3]{Michael Saunders}
\author[4]{Andrea Walther}
\author[1]{Olaf Schenk}
\affil[1]{Universit\`a della Svizzera italiana}
\affil[2]{University of Ioannina}
\affil[3]{Stanford University}
\affil[4]{University of Paderborn}



\abstract{
  Template metaprogramming is a popular technique for
  implementing compile time mechanisms for numerical computing.  We
  demonstrate how expression templates can be used for compile time
  symbolic differentiation of algebraic expressions in \Cpp{} computer
  programs.  Given a positive integer $N$ and an algebraic function of
  multiple variables, the compiler generates executable code for the
  $N$th partial derivatives of the function.  Compile-time
  simplification of the derivative expressions is achieved using
  recursive templates.  A detailed analysis indicates that current
  \Cpp{} compiler technology is already sufficient for practical use
  of our results, and highlights a number of issues where further
  improvements may be desirable.
}

\renewcommand{\arraystretch}{1.2}

\newcommand\eq{\begin {eqnarray}}
\newcommand\eeq{\end {eqnarray}}
\newcommand\ds {\displaystyle}
\newcommand\scr {\scriptscriptstyle}

\def\NPT{{\em NPT}}
\def\DegK{{\rm K}}
\def\Angst{{\rm\AA}}
\def\rad{{\rm rad}}



\date{}

\maketitle
\section{Introduction}    
\label{sec:intro}

Methods employed for the solution of scientific and engineering
problems often require the evaluation of first or higher-order
derivatives of algebraic functions.  Gradient methods for
optimization, Newton's method for the solution of nonlinear systems,
numerical solution of stiff ordinary differential equations, stability
analysis: these are examples of the major importance of derivative
evaluation.  Computing derivatives quickly and accurately improves
both the efficiency and robustness of such numerical algorithms.
Automatic differentiation tools are therefore increasingly available
for the important programming languages; see \url{www.autodiff.org}.

There are three well established ways to compute
derivatives:

{\it Numerical derivatives}. These use finite-difference
approximations \cite{NR}. They avoid the difficulty of very long
exact expressions, but introduce truncation errors and this usually
affects the accuracy of further computations. An intriguing alternative 
that in contrast to finite-difference approximations obtains the exact 
derivatives up to machine precision, and it is easy to implement provided that
the function under consideration is analytic, is the complex-step approach
\cite{Martins:2003:CSD}. 

{\it Automatic differentiation (AD)}. This is a way to find the
  derivative of an expression without finding an expression for the
  derivative.  Specifically, in a ``computing environment'' using AD tools,
  one can obtain a numerical value for $f'(x)$ by providing an
  expression for $f(x)$. The derivative computation is accurate to
  machine precision.  A good introduction to the methods for
  implementing AD and the concepts underlying the method can be found
  in \cite{GriewankSIAM2000,GriewankSIAM1991,Griewank01AD89,Griewank01,BischofADIFOR01,Rall}.
  Several other software packages implement AD approaches. Given a set of Fortran
subroutines for evaluating a function $f$, 
ADIFOR~\cite{BischofADIFOR01,ADIFORhtml} produces Fortran 77
subroutines for computing the first derivatives of the function.
Upgrades and extensions in other high-level programming 
languages such as C and \Cpp{} now exist \cite{ADIChtml}.
\FADBADpp{} and ADOL-C are \Cpp{} libraries that combine the two
basic ways (forward/backward) of applying the chain rule~\cite{Walther2012,ADOLCBook,Griewank,FadBadhtml,FadBadreport}. 
Aubert et al.\ implement automatic
differentiation of \Cpp{} computer programs in forward mode using
operator overloading and expression templates \cite{Aubert01}. These libraries
have demonstrated the ability to perform sensitivity analysis by marginally 
modifying the source of the computer program, replacing the double type to the type
implemented in the provided library, and simply linking with the library.
The implementation of the reverse mode using expression
templates forms a different task because the program flow has
to be reversed in this case. The implementation of AD is straightforward in the environment of the
object-oriented high-level language \Cpp{} with operator overloading
and expression templates \cite{StroustrupC,BartonNackmanC,MeyersC}.

{\it Symbolic derivatives}.
These are obtained by hand or from one of
  the symbolic differentiation packages such as Maple, Mathematica, or
  {\sc Matlab}.  Hand-coding is increasingly difficult and
  error-prone as the function complexity increases.
  Symbolic differentiation packages can obtain expressions for the
  derivatives using the rules of calculus in a more or less mechanical
  way. Given a string describing a function, they provide exact
  derivatives by expressing them in terms of intermediate
  variables. This method provides a formula for the first derivative,
  which can be further differentiated if derivatives of higher order
  are desired. Since the formulae for the derivatives are exact, the
  approach does not introduce any truncation errors, unlike the other
  differentiation methods.

In this work we present a new way of obtaining partial derivatives of
arbitrary order for multivariate functions, in a way that exhibits
optimal runtime performance.  This is achieved by exploiting the
\Cpp{} Expression Templates mechanism described next.

\section{\protect\Cpp{} templates}    
\label{sec:templatesc++}

Templates were introduced in \Cpp{} to allow type-safe containers.  In
the early days, templates
served mostly as a means of generalizing software components so they
could be easily reused in a variety of situations. Templates' ability to allow
generalization without sacrificing efficiency made them an integral
tool of generic programming. Eventually it was discovered by Unruh
\cite{Unruh}, almost by accident, that the \Cpp{} template mechanism
provides a rich facility for native language metaprogramming: the
creation of programs that execute inside \Cpp{} compilers and that
stop running when compilation is complete. Today, the power
of templates is fully unleashed
\cite{Alexandrescubook,VanderJosBook,BartonNackmanC}, and template
metaprogramming has been extensively investigated by several authors
\cite{VeldhuizenBlitz,Abrahamsbook,Boosthttp}.

Moreover, the combination of classical \Cpp{} operator overloading
with template metaprogramming ideas has resulted in a very promising
technique, \emph{expression templates}, that has found numerous
applications in scientific computing. In \cite{Blaze,VanderJosBook} the
authors explain how expression templates can be used to construct an
efficient library for matrix algebra avoiding  introducing runtime
temporary matrix objects, with have an adverse performance and memory
management effect.  In contrast, the combination of expression templates 
and sophisticated optimisation techniques build in the \Cpp{} compilers
used for the generation of the executable code, can efficiently 
eliminate temporary objects in many situations and thus do not suffer 
from any performance or memory issues inherent in the creation and destruction of temporaries.
The object-oriented interface can be preserved without sacrificing
efficiency.  Veldhuizen \cite{VeldhuizenBlitz} presents a \Cpp{} class
library for scientific computing that provides performance on a par
with Fortran 77/90.  Advanced language features are maintained while
utilization of highly sophisticated template techniques ensures no
performance penalty at all. If templates are used appropriately,
optimizations such as loop fusion, unrolling, tiling, and algorithm
specialization can be performed automatically at compile time.

In \cite{Aubert01}, the authors use expression templates to handle
automatic differentiation of multivariate function objects and apply
this technique to a control flow problem. Their approach, being the
first application of expression templates in the area of automatic
differentiation, lacks some important features.  In particular, the
partial derivative of a multivariate function object provided by the
user is not constructed at compile time.  Instead, its value is
calculated at runtime from the derivatives of all sub-expressions
that comprise the main expression of the function to be
differentiated. This approach is suboptimal because trivial
calculations (like multiplications by one or zero) are not eliminated
and performance penalties may occur, especially if the derivative has
to be evaluated at a large number of points. Applications of
expression templates for the efficient calculation of derivatives and
Jacobians have been reported by Younis~\cite{Younis:2007}. These 
techniques were adopted by Kourounis et al.~\cite{Kourounis:2014}
for the evaluation of the individual derivatives needed by the
discrete adjoint formulation in applications involving the control
and optimization of compositional flow in porous media. A recent
application of expression templates can be found in~\cite{Hogan:2014},
where a new operator-overloading method is presented that provides 
a compile-time representation of mathematical expressions as a computational 
graph that can be efficiently traversed in each direction. However,
the expressions obtained this way cannot be further differentiated
and the user can only expect first order derivatives.

In this paper we try to improve the ideas in \cite{Aubert01} and \cite{Nehmeier} and to extend
them in a number of ways. We show that partial derivatives of any order can be
constructed upon request at compile time, as function objects themselves.
Unlike the approach presented by Nehmeier~\cite{Nehmeier}
we enhance our approach by introducing simplification rules performed in
compile time. Without claiming completeness, we demonstrate
how template metaprogramming techniques could be employed to simplify the
resulting expression for the partial derivatives during compilation. Trivial
calculations are thus eliminated. Further algebraic simplifications, such 
as cancellation of common terms, are also performed where possible. 

We refer to our approach by the name CoDET (Compile-time
Differentiation using Expression Templates).  After introducing the
key concepts, we describe experiments with a number of different
\Cpp{} compilers to benchmark the compile time and scalability of
CoDET over large expressions, while assessing the quality of the
generated code.  Several examples demonstrate that the execution cost
of the partial derivative constructed by CoDET is identical to that of
a hand-coded version.  Along the way, we identify a number of
compiler-related issues and optimizations that affect these costs and
suggest compiler features and further enhancements beneficial for our
approach.

\section{Multivariate expression definition}    
\label{sec:model}

The purpose of this section is to expose and analyze all the classes that take
part in the implementation of CoDET's multivariate expressions.

\begin{figure}[!tb]   
\begin{center}
  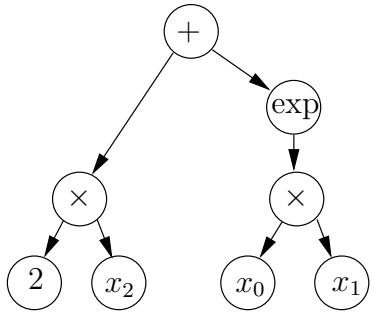
\end{center}
\caption{Tree for evaluating the expression $2 x_2 + e^{x_0 x_1}$.}
\label{fig:est1}
\end{figure}

The CoDET framework is inspired by the Expression Templates of
\cite{VeldhuizenExpTemp95}.  Each expression is modeled by an
\emph{expression syntax tree (EST)}, whose leaves are either numeric
constants or independent variables, and whose internal nodes
correspond to functions (unary, binary or $n$-ary) or operators
(arithmetic, logical, etc.)\ on the sub\-expressions of the
corresponding sub\-trees. For example, the expression $2 x_2 + e^{x_0
  x_1}$ can be modeled by the EST in Figure~\ref{fig:est1}.

In order to perform compile time symbolic differentiation, we use the \Cpp{}
type system to encode algebraic expressions in a manner isomorphic to ESTs.
A \Cpp{} class type corresponds to each node of an expression syntax tree.
Leaf nodes of the EST are either variables or constants. Different variables
correspond to different classes, as do different constants. Unary internal
nodes of the tree correspond to analytical functions (such as $\sin$, $\exp$,
$\log$, etc.)\ or to the negation operator.  Binary internal nodes correspond
to binary arithmetic operations.

\subsection{Encoding Multivariate Expressions as Types}
\label{ssec:encexpr}
\Cpp{} template class instantiations are not a notationally appropriate form
for denoting algebraic expressions. Thus, our framework strives to support the
implicit declaration of these template instantiations by employing function
and operator overloading.

For instance, assume that we are working in $\mathbb{R}^3$ and we want to
declare a function of three independent variables: $f(\mathbf{x}) = f(x_0,
x_1, x_2) = 2x_2 + e^{x_0 x_1}$.  Using the classes \lstinline+Variable<int N>+
and \lstinline+Real<int L,int R,int Ex>+ introduced
in \S\ref{ssec:Variables} and \S\ref{ssec:real-constants} below,
our framework allows us to write
\begin{lstlisting} 
Variable<0> x0;
Variable<1> x1;
Variable<2> x2;
Real<2,0,1> _2_;   // stands for 2 = 0.20e1
typedef decltype( _2_ * x2 + exp(x0 * x1) ) fType;
\end{lstlisting}
in which we declare class type \lstinline+fType+ to correspond to the
expression $2x_2+e^{x_0 x_1}$, and the code inside \typeof{} closely
matches the algebraic form. 
Once an expression type is defined, it can be instantiated and its instances
can be used to evaluate the expression:
\begin{lstlisting}%[numbers=left]
fType f;
double x[] = { 1.0, 2.5, 3.14 };
cout << f(x) << "\n"; // outputs the value of 2*3.14+exp(-2.5)
\end{lstlisting}

The \auto{} keyword that is included in the \Cpp{11} standard,
specifies that the type of the variable that is being declared will be 
automatically deduced from its initializer.
The use of the \auto{} keyword would allow simpler
code; for example:

\begin{lstlisting}
Variable<0> x0;
Variable<1> x1;
Variable<2> x2;
Real<2,0,1> _2_;

auto f =  _2_ * x2 + exp(x0 * x1);
double x[] = { 1.0, 2.5, 3.14 };
cout << f(x) << "\n"; // outputs the value of 2*3.14+exp(-2.5)
\end{lstlisting}

\subsection{Arithmetic operators and analytical expressions}
\label{ssec:operators}

Arithmetic operators and analytical expressions are supported by class
templates that are parameterized by the types of their subexpressions.
Let us examine binary arithmetic operators first.  Following
\cite{VeldhuizenBlitz}, we start by introducing the descriptor
(non-templatized) classes \lstinline{Add}, \lstinline{Sub}, \lstinline{Mul},
\lstinline{Div}. We list the source code only for the definition of
class \lstinline{Add}. The definitions of the remaining classes follow in a
similar manner.

\begin{lstlisting}
class Add
{
public:
    inline static double apply(double a, double b)
    { return a+b; }
};
\end{lstlisting}

The above classes are used to parameterize class template \lstinline{BinaryOp},
the class that implements arithmetic binary operations between types:
\begin{lstlisting}
template<typename L,typename R,typename Op>
class BinaryOp
{
public:
    L left_;
    R right_;
    inline double operator()(const double* x) const
    { return Op::apply(left_(x), right_(x)); }
};
\end{lstlisting}

To construct expressions easily, we provide templatized versions of \Cpp{}
arithmetic operators. We show only the definition for \lstinline{operator+}.
Overloaded versions of the remaining operators are defined similarly.
\begin{lstlisting}[title={\itshape \lstinline{L}$+$\lstinline{R} $\rightarrow$ \lstinline!BinaryOp<L,R,Add>!}]
template<typename L,typename R> inline
BinaryOp<L,R,Add> operator+(const L& rleft, const R& rright)
{ return BinaryOp<L,R,Add>(rleft, rright); }
\end{lstlisting}

Analytical functions are unary operators, and they can be defined in a
manner similar to the definition of binary operators. However, we have
implemented them in a more direct manner in order to simplify coding.
The definition of the node for function \lstinline+exp+ follows,
together with an overloaded function template for easy expression
construction:
\begin{lstlisting}
template<typename T>
class MathExp
{
public:
    T expr_;
    inline double operator()(const double* x) const 
    { return exp( expr_(x) ); }
};
\end{lstlisting}

\begin{lstlisting}[title={\itshape \lstinline+exp(T)+ $\rightarrow$ \lstinline+MathExp<T>+}]
template<typename T> 
inline MathExp<T> exp(const T& rfexpr)
{ return MathExp<T>(rfexpr); }
\end{lstlisting}

\subsection{Variables}
\label{ssec:Variables}

The CoDET framework supports an arbitrary number of independent variables.
Each of these variables corresponds to an instance of class template
\lstinline+Variable<int>+.

The definition of the class template \lstinline+Variable+
is quite straightforward:

\begin{lstlisting}
template<int varID>
class Variable
{
public:
    double operator()(const double* x) const 
    { return x[varID]; }
};
\end{lstlisting}

\subsection{Integer and real constants}
\label{ssec:real-constants}

CoDET provides two approaches for implementing constants,
again using template instances:

{\it The {\rm\approachint} approach.} This exploits the class
  \lstinline+Integer<int Value>+.  Although the class is restricted
  to integer constants, its adoption leads to remarkable compilation
  time savings.  (We would prefer to use the template class
  \lstinline+Real<double Value>+, which is supported by the D language
  but not by \Cpp. We strongly believe that it would share the same
  features as the currently available integer-only version.)
  
{\it The {\rm\approachreal} approach.} This combines class
  \lstinline+Real<int L,int R,int Ex>+, which can represent any real
  number, with class \lstinline+Constant<typename T>+, which wraps every
  arithmetic operation between constants and thereby allows specific
  floating-point optimizations.
Both approaches are used in \S\ref{ssec:simplifications} and
discussed further in \S\ref{ssec:template-constants}.
Three new classes are needed, as now described.

\subsubsection{The class Integer}
\label{sssec:constants-integer}

To support the frequent case of integer-valued constants (of type
\lstinline{double}), we provide the following class for integer constants:

\begin{lstlisting}
template<int Value> 
class Integer
{
public:
    double operator()(const double* x) 
    { return double(Value); }
};
\end{lstlisting}
The most important feature of this class is that it allows arithmetic
operations between integers to be performed during the compilation process.
To be more precise, let us consider the function $f(x) = e^{3x}$. Its fourth
derivative will be $d^4f(x)/dx^4 = 3 \cdot 3 \cdot3 \cdot 3 e^{3x}$. With
appropriate simplification rules, all intermediate multiplications can be
performed during compilation to give the formula
$d^4f(x)/dx^4 = 81 e^{3x}$.  However, the range of integers that
can be represented is limited to that of the integer type provided by the
\Cpp{} language.  (Again, we would like a class where the template
parameter is of type \lstinline{double} and not \lstinline{int}.)
We proceed by introducing a class that does not suffer from such limitations.

\subsubsection{The class Real}
\label{sssec:constants-real}

This class uses three integer parameters to
compose the value of the double precision constant represented by the class:
\begin{lstlisting}
template<int L,int R,int Ex>
class Real
{
private:
    double value_;

public:
    Real()
    {
      std::ostringstream strout;
      strout << "0." << L << R << "e" << Ex;
      value_ = double( strtod(name.c_str(), NULL) );
    }

    inline double operator()(const double* x) const 
    { return value_; }
};
\end{lstlisting}

The first two template parameters are integers that represent when
written sequentially (\lstinline+LR+) the decimal digits of our
constant. The third template parameter \lstinline+Ex+ stands for the
exponent. For example, we can write the constant 1234.56789 as
\lstinline+Real<1234,56789,4>+.  Inside the default constructor this
will be converted to the literal value ``0.123456789e4'', and later
this literal will be converted to an arithmetic value of type
\lstinline+double+ and stored in the private member \lstinline+value_+. The
\lstinline+operator()+ always returns the same value, independently of the
\lstinline+double+ pointer passed to it. Unfortunately, this class does not
allow simplification of arithmetic operations between real constants.
A work-around is presented next.

\subsubsection{The class Constant}
\label{sssec:constants-constant}

Suppose we would like to differentiate repeatedly the function
$f(x)=e^{2.3x}$.  If the constant $2.3$ is implemented using the class
\lstinline+Real+ as \lstinline+Real<2,3,1>+, then the $N$th derivative
$\frac{d^Nf(x)}{dx^N} = (2.3)^{N} e^{2.3x}$ would require $N-1$
multiplications every time it is called. It would be very convenient if
constants like this could be evaluated once and for all during the
construction of the function object, as their values are independent of $x$.
For this purpose we introduce the following class, which is a wrapper of all
such operations between constants:
\begin{lstlisting}
template<typename T>
class Constant
{
private:
    T expr_;
    double value_;

public:    
    Constant()
        : expr_(T())
        { value_ = expr_(0); }  
    double operator()(const real* x) const
        { return value_; }
};
\end{lstlisting}

Appropriate simplification rules, performed at compile time, ensure that this
class wraps every arithmetic operation between constants. 
In this way, during 
construction of the function object, the default constructor of this class
calculates the constant expression represented by the type of the object
\lstinline+expr_+ and assigns its value to the private member
\lstinline+value_+ of type \lstinline+double+.  Subsequent calls to the
\lstinline+operator()+ of this class will not involve any intermediate
calculations such as $2.3\cdot2.3\cdot2.3\cdot\ldots2.3$ because the result
has been computed once and for all in the default constructor.

For example, to be able to wrap successive multiplications of
real constants like $2.3\cdot2.3\cdot2.3\cdot\ldots2.3$, we need to provide
a set of simplification rules using the following class \lstinline+Squeezer+:

\begin{lstlisting}
template<typename T>
class Squeezer
{
public:
    typedef T squeezedType;
};
\end{lstlisting}

This class operates recursively on its type argument in order to simplify it
as much as possible, guided by appropriate simplification rules.  The
resulting simplified version of type \lstinline+T+ can be obtained as the
nested type name \lstinline+squeezedType+.  The simplification rules are
provided as explicit template specialization of the class
\lstinline+Squeezer+.  In explaining the role of the rules needed for
our case, we represent the class \lstinline+BinaryOp<A,B,Op>+ by
\lstinline+A+$\circledast$\lstinline+B+,
where $\circledast$ may be one of the four common binary
arithmetic operators $+,-,*,/$.

The following rule ensures that we will always have one \lstinline+Constant+
wrapping every arithmetic operation between objects of type \lstinline+Constant+:
\begin{lstlisting}[title={\itshape \lstinline+Constant<A>+$\circledast$\lstinline+Constant<B>+
    $\rightarrow$ \lstinline+Constant<A+$\circledast$\lstinline+B>+}]
template<typename A,typename B,typename Op>
class Squeezer<BinaryOp<Constant<A>,Constant<B>,Op> >
{
public:
    typedef Constant<BinaryOp<A,B,Op> > squeezedType;
};
\end{lstlisting}
Then we need to ensure that objects of type \lstinline+Constant+ enclosed in
class \lstinline+BinaryOp+ will be visible from constants operating on that 
class.  This is achieved for multiplication by the following rule:
\begin{lstlisting}[title={\itshape \lstinline+Constant<A>+$*($\lstinline+Constant<B>+$*$\lstinline+C+$)
  \rightarrow$ \lstinline+Constant<A+$*$\lstinline+B>+$*$\lstinline+C+}]
template<typename A,typename B,typename C>
class Squeezer<BinaryOp<Constant<A>,BinaryOp<Constant<B>,C,Mul>,Mul> >
{
public:
    typedef BinaryOp<Constant<BinaryOp<A,B,Mul> >,
                     typename Squeezer<C>::squeezedType,
                     Mul> squeezedType;
};
\end{lstlisting}
We only need to specify the previous rule for constants appearing as left
operands, because appropriate overloading of the binary operator $*$
ensures that this is always the case. The binary operator of multiplication
has to be overloaded as follows:
\begin{lstlisting}[title={\itshape\lstinline+T+$*$\lstinline+Constant<Real<L,R,Ex>>+
  $\rightarrow$ \lstinline+BinaryOp<Constant<Real<L,R,Ex>>,T,Mul>+}]
template<typename T,int L,int R,int Ex> inline 
BinaryOp<Constant<Real<L,R,Ex> >,T,Mul> operator*(const T& rleft,
                                                        const Real<L,R,Ex>& rc)
{
    typedef BinaryOp<Constant<Real<L,R,Ex> >,T,Mul> exprT;
    return exprT(Constant<Real<L,R,Ex> >(), rleft);
}
\end{lstlisting}
\begin{lstlisting}[title={\itshape \lstinline+Constant<Real<L,R,Ex>>+$*$\lstinline+T+
  $\rightarrow$ \lstinline+BinaryOp<Constant<Real<L,R,Ex>>,T,Mul>+}]
template<typename T,int L,int R,int Ex> inline 
BinaryOp<Constant<Real<L,R,Ex> >,T,Mul> operator*(const Real<L,R,Ex>& rc,
                                                        const T& rright)
{
    typedef BinaryOp<Constant<Real<L,R,Ex> >,T,Mul> exprT;
    return exprT(Constant<Real<L,R,Ex> >(), rright);
}
\end{lstlisting}

\section{Compile-time partial derivatives}    
\label{sec:diff}

We now turn our attention to partial differentiation of expressions
represented as ESTs, where the partial derivatives are also
represented as ESTs.  For the example of
\S\ref{ssec:encexpr}, our framework computes the partial derivatives
of \lstinline{fType} as follows:
\begin{lstlisting}
Der<0, fType>::derType df_dx0;
Der<1, fType>::derType df_dx1;
Der<2, fType>::derType df_dx2;
\end{lstlisting}

Here, class template \lstinline{Der} is parameterized by two types: the type of
the differentiation variable and the type of the expression to be
differentiated. The derivative of the expression is then obtained as the
nested type name \lstinline{derType}.  The basic technique is extensive use of
specializations of class template \lstinline{Der}, where each specialization
corresponds to a particular node type of the ESTs.  Differentiation then
proceeds recursively down the input EST, generating the EST of the derivative.

However, a naive implementation of differentiation as above is
inefficient and non-scalable because the resulting ESTs would grow
prohibitively large (in the worst case exponentially) compared to the
original EST, mostly because of a large number of trivial operations
(addition of zero, multiplication by zero or one, etc.).  If not
handled well, this explosion in size of the expression trees will
affect both compilation (which will become unacceptably slow and
require excessive memory) and runtime, because evaluation of ESTs
takes time directly proportional to their size.

Thus, it is imperative that intermediate expressions produced during
differentiation be simplified algebraically. Conceptually, such simplification
could be carried out as a postprocessing step on the ESTs of the derivatives.
This approach would yield efficient runtime expression evaluation of the
generated ESTs, but would increase the effort at compile time, reducing the
scalability of our technique to expressions of only modest size.

Our approach is to perform simplifications interleaved with the
differentiation steps, and to limit our simplification patterns to a carefully
selected set of rules.  Simplification patterns are
defined by appropriate specializations of class template \lstinline{Squeezer},
which was introduced in \S\ref{sssec:constants-constant}.
We now outline our overall approach in more detail.

\subsection{Differentiating constants and variables}

We introduce two convenient constant definitions:
\begin{lstlisting}
typedef Real<0,0,0> Zero;
typedef Real<1,0,1> One;
\end{lstlisting}

The differentiation rules for the leaves of ESTs, namely variables
and constants, are non-recursive. The rule for differentiating constants 
is the simplest:
\begin{lstlisting}
template <int N,int L,int R,int Ex>
class Der<N, Real<L,R,Ex> > 
{
public: 
    typedef Zero derivType; 
};
\end{lstlisting}
A similar rule exists for classes of type \lstinline{Integer}.

For variables, we use two specializations. The first one corresponds to the
rule $\partial x_N/\partial x_N = 1$, whereas the second corresponds to 
$\partial x_M/ \partial x_N = 0$.
\begin{lstlisting}
template<int N>    
class Der<Variable<N>,Variable<N> > 
{ 
public: 
    typename One derivType;
};
\end{lstlisting}
\begin{lstlisting}
template<int N,int M>
class Der<Variable<N>,Variable<M> > 
{ 
public: 
    typename Zero derivType;
};
\end{lstlisting}

\subsection{Differentiation of arithmetic and analytical expressions}

Differentiation rules for internal nodes of ESTs are recursive. Let us
consider the basic product rule and its implementation:
\begin{lstlisting}[title={\small$\displaystyle{
  \frac{\partial (L*R)}{\partial x} \rightarrow 
  \frac{\partial L}{\partial x} \mathrel* R + 
  L \mathrel*\frac{\partial R}{\partial x}}$}]
template<int N,typename L,typename R>
class Der<N,BinaryOp<L,R,Mul> >
{
    typedef typename Der<N,L>::derivType _dL;
    typedef typename Der<N,R>::derivType _dR;
public:
    typedef typename BinaryOp<BinaryOp<_dL,R,Mul>, 
                                 BinaryOp<L,_dR,Mul>, 
                                 Add> derivType;
};
\end{lstlisting}
To achieve simplification, differentiation rules are interlaced with the
\lstinline{Squeezer} simplification pattern.
For example, let $\Squeeze(F)$ represent the simplified form of $F$.
The product rule then becomes the following:
\begin{lstlisting}[title={\small$\displaystyle{
    \frac{\partial (L*R)}{\partial x} \rightarrow
    \Squeeze\biggl(
      \Squeeze\bigl(\frac{\partial L}{\partial x}\bigr)
        \mathrel* R + L \mathrel*
      \Squeeze\bigl(\frac{\partial R}{\partial x}\bigr)
    \biggr)}$}]
template<int N,typename L,typename R>
class Der<N,BinaryOp<L,R,Mul> >
{
    typedef typename 
        Squeezer<typename Der<N,L>::derivType>::squeezedType _dL;
    typedef typename 
        Squeezer<typename Der<N,R>::derivType>::squeezedType _dR;
public:
    typedef typename 
        Squeezer<BinaryOp<BinaryOp<_dL,R,Mul>, 
                            BinaryOp<L,_dR,Mul>, 
                            Add> >::squeezedType  derivType;
};
\end{lstlisting}
Similarly for the exponential, the rule and code without
simplifications are as follows:
\begin{lstlisting}[title={\small$\displaystyle{
  \frac{\partial \exp(F)}{\partial x} \rightarrow
  \exp(F)\mathrel*\frac{\partial F}{\partial x}}$}]
template<int N,typename F>
class Der<N,MathExp<F> >
{
    typedef typename Der<N,F>::derivType _dF;
public:
    typedef typename BinaryOp<MathExp<F>,_dF,Mul> derType;
};
\end{lstlisting}
However, when we wish the compiler to perform simplifications during
compilation, the above rule and code have to be modified:
\begin{lstlisting}[title={\small$\displaystyle{
  \frac{\partial \exp(F)}{\partial x} \rightarrow
 \Squeeze\Bigl( \exp(F)\mathrel*
 \Squeeze( \frac{\partial F}{\partial x}) \Bigr)}$}]
template<int N,typename F>
class Der<N,MathExp<F> >
{
    typedef typename 
        Squeezer<typename Der<N,F>::derivType>::squeezedType _dF;
public:
    typedef typename 
        Squeezer<BinaryOp<MathExp<F>,_dF,Mul> >::squeezedType derType;
};
\end{lstlisting}
Other arithmetic operators and analytic functions 
are handled in the same spirit.

\subsection{Expression simplifications}

To appreciate the effect of simplification on the size of the
derivative EST of an expression, consider the partial derivative with
respect to $x_1$ of $2\cdot (x_1\cdot \exp(x_2))$ (computed by the
product and exponential rules). Without simplifications, one would get
\[ 
  0\cdot (x_1 \cdot \exp(x_2)) + 
  2\cdot [1\cdot \exp(x_2) + x_1\cdot (\exp(x_2) \cdot 0) ]
\]
instead of the relatively simple expression $2\exp(x_2)$ (an EST of 21
nodes instead of just 4).  Evaluation of the unsimplified formula
could have much higher runtime than for the simplified one.
Additionally, the compile time differentiation of the original formula
would flood the compiler's symbol tables with a plethora of trivial
types, increasing compilation time and memory use.

Algebraic simplification is an old and broadly studied subject of
symbolic computation. Simplification rewrites a given EST as a new EST
that is in some sense simpler.  Several projects on template
metaprogramming include expression simplifiers; e.g., Schupp et~al.\
\cite{udsimplification} describe a user extensible simplification
framework for expression templates over abstract data types.

The rules of a simplifier must be chosen carefully; a limited set
of rules might miss significant simplification opportunities, but
a very extensive set might introduce significant compilation overhead and
result in dubious simplicity---for example, which expression is ``simpler'':
$x\cdot x - y\cdot y$ or $(x-y)\cdot (x+y)$?
We have tested and propose the rules shown in Table~\ref{tab:simprules}.
\begin{table}[tb]  
\caption{Simplifier rewrite rules. $x, y, z$ represent arbitrary formulae;
$n, m$ represent integer-valued constants.\label{tab:simprules}}
\begin{flalign*}
   & & x \pm 0           &\to x      &   (x y)/(x z) &\to y/z & &
\\ & & x \cdot 0         &\to 0      &   (x y)/x     &\to y   & &
\\ & & x\cdot 1          &\to 1      &   x/(x y)     &\to 1/y & &
\\ & & n\cdot (m\cdot x) &\to (nm) x &   x/x         &\to 1   & &
\\ & & -(x-y)            &\to y-x    &   0/x         &\to 0   & &
\\ & & -(-x)             &\to x      &   x/1         &\to x   & &
\\ & & x+(-y)            &\to x-y    &   1/(x/y)     &\to y/x & &
\\ & &                   &           &   x(1/y)      &\to x/y & &
\end{flalign*}
\end{table}
Note that a single rewrite rule may need to be implemented by several template
specializations.  The following rules simplify addition of zero to some
variable or expression. The first template specialization implements the rule
$x+0\rightarrow x$ while the second implements $0+x\rightarrow x$. The first
two specializations cater to the commutativity of addition.  The last
specialization is needed by the compiler to resolve the ambiguity between the
first two when both template parameters are objects of type \lstinline+Zero+.
Simplification rules for multiplication by zero or one are defined in a
similar way.
\begin{lstlisting}[title={\small$x+0\rightarrow x$}]
template<typename T>  
class Squeezer<BinaryOp<T,Zero,Add> >
{ 
public: 
    typedef T squeezedType; 
};
\end{lstlisting}
\begin{lstlisting}[title={\small$0+x \rightarrow x$}]
template<typename T>  
class Squeezer<BinaryOp<Zero,T,Add> >
{ 
public: 
    typedef T squeezedType; 
};
\end{lstlisting}
\begin{lstlisting}[title={\small$0+0 \rightarrow 0$}]
template<>  
class Squeezer<BinaryOp<Zero,Zero,Add> >
{ 
public:  
    typedef Zero squeezedType; 
};
\end{lstlisting}

In studying Table~\ref{tab:simprules}, one might notice certain discrepancies.
For example, there is a rule $x/(xy) \to 1/y$ but not the equivalent addition
rule $x-(x+y) \to -y$. The reason is that while the first rule applies to a
number of expressions the user is likely to write (e.g., differentiating the
expression tree $x/2$ with respect to $x$), the second would apply in unlikely
formulae only.  In our design 
we have chosen to keep a rather minimal set of
simplifiers, as our purpose is not to develop a complete compile time symbolic
differentiation package but rather to illustrate the idea and motivate further
developments.

\subsection{Higher-order derivatives}

A straightforward way to obtain higher-order partial derivatives is by
sequential differentiation. As a form of programming convenience, we
provide a recursive class template, \lstinline{DerN}, for obtaining
the $N$th derivative of an expression $F(x)$ with respect to variable
$x_M$:
\begin{lstlisting}[title={\small$\displaystyle{
    \frac{\partial^N F}{\partial x_M^N} \rightarrow
    \frac{\partial^{N-1} \left( \partial F / \partial x_M \right )}
         {\partial x_M^{N-1}}}$}]
template<int N,int M,typename F>
class DerN 
{
public:
    typedef typename 
      DerN<N-1,M,typename Der<M,F>::derivType>::derivType derivType;
};
\end{lstlisting}
The following specialization is needed to end the recursion:
\begin{lstlisting}[title={\small$\displaystyle{
   \frac{\partial^1 F}{\partial x_M^1} \rightarrow
   \frac{\partial F}  {\partial x_M}}$}]
template<int N,typename F>
class DerN<1,M,F> 
{
public: 
    typedef typename Der<M,F>::derType derType;
};

// example:  f(x):=  d^2(exp(x*x)) / dx^2
Variable<0> x;
typedef DerN<2,0,decltype(exp(x*x))>::derivType f;
\end{lstlisting}

\begin{remark}
  Our methodology allows evaluation of any partial derivative of
  arbitrarily high order. For expressions with derivative formulas
  that remain bounded independently of the differentiation order, and
  provided that the set of simplification rules implemented can handle
  every possible case, our approach generates a template expression
  with number of terms also bounded independently of the
  differentiation order.  In the general case, however, the highest
  order of differentiation may be restricted by several factors. The
  time needed for the differentiation grows linearly with the number
  of terms that are inlined. For problems where the formulas of the
  partial derivatives grow exponentially in size with respect to the
  differentiation order, the same is observed with compilation time
  and memory.
\end{remark}


\section{Empirical evaluation}     
\label{sec:empiricalEvaluation}

The effectiveness of the CoDET approach is investigated in this
section through several test cases. To verify that our results are
independent of CPU architectures we used the following
platforms running the same 64-bit Linux distribution:
\begin{enumerate}
\item \textbf{Intel Xeon CPU E5-2670}
      \\ 2.60 GHz, 20480 KB L3 cache
\end{enumerate}
All test cases benchmarked four different \Cpp{} compilers.
The compilers and their compilation flags follow:
\begin{enumerate}
\item \textbf{GNU (GCC) 4.8.2}
      \\ \textit{g\pp-4.8.2 -static -O3}
\item \textbf{Intel icpc (ICC) 14.0.0 20130728}
      \\ \textit{icpc -O3}
\item \textbf{Sun 5.9 Linux\_i386 Patch 124865-01 2007/07/30}
      \\ \textit{CC -xO5 -features=extensions -m64}
\item \textbf{Portland Group pgCC 14.4-0 64-bit}
      \\ \textit{pgCC -O3 --gnu}
\end{enumerate}


For each test we present both compilation time (for the
compilers to generate the executable file) and runtime performance of
the benchmarked function \lstinline{f(x)} for executing the following loop:
\begin{lstlisting}
double sum,  x[1];
x[0] = 0.0;
sum = 0.0;
for (i = 0; i < numberOfIterations; i++)
{
    x[0] -= 0.1;
    sum += f(x);
}
\end{lstlisting}

In all examples we set $\hbox{\lstinline{numberOfIterations}} = 10^7$.
The loop overhead varies from $0$ to $100$ milliseconds at most and is
subtracted from the running time of the above loop. Thus, the
runtimes measured here essentially correspond to the total time
needed for the function calls.

The code that instructs the compiler to generate the $N$th derivative
for the simple univariate function $f(x) = e^x + e^{2x} + e^{3x}$ is
as follows:
\begin{lstlisting}
Variable<0> X;
typedef decltype(            
           exp( Constant<1>()*X ) 
         + exp( Constant<2>()*X ) 
         + exp( Constant<3>()*X ) ) fType;

// our functional f = e^x + e^(2x) + e^(3x)
fType f = exp( Constant<1>()*X ) 
         + exp( Constant<2>()*X )  
         + exp( Constant<3>()*X );

// and its Nth derivative
const int N = 1;
typedef DerN<N,0,fType>::derType dfType;
dfType dNf_dxN;
\end{lstlisting}
By changing the value of the constant variable $N$ we obtain the
desired order of the derivative. The formula for $f$ changes according
to the benchmarked case.  Note that the support of
the \Cpp{} type \auto{}, currently provided by the \Cpp{11} standard, allows 
us simpler code:
\begin{lstlisting}
Variable<0> X;

auto f = exp( Constant<1>()*X ) 
        + exp( Constant<2>()*X )  
        + exp( Constant<3>()*X );

const int N = 1;
auto dNf_dxN = derivative<N>(f);
\end{lstlisting}
where the definition for the function \lstinline+derivative()+ follows:
\begin{lstlisting}
template<int N, typename T> auto derivative(T f)
{
  typedef typename DerN<N, 0, T>::derType dfType;
  return dfType();
}
\end{lstlisting}
In the same spirit one can introduce a similar function taking two integer
template arguments to allow simpler code for partial derivatives.

\subsection{The need for compile-time simplifications}
\label{ssec:simplifications}

Simplifications are an essential ingredient of every symbolic
package. In the present study, compilers without them would not be
able to cope with higher-order derivatives, as intermediate
expressions would grow exponentially and the generated code would
perform poorly. This is demonstrated in what follows.

\subsubsection{Simplifications disabled} 

Consider again the function $f(x) = e^x + e^{2x} + e^{3x}$.
With simplifications disabled in our code, we examine the compile time
for each compiler and the runtime of the
compiler-generated derivatives up to order six.  Constants were
implemented by the \approachint{} approach of
\S\ref{sssec:constants-integer}.  The results obtained using the
\approachreal{} approach are similar and do not provide further
insights.


\begin{figure}[tb]   
\begin{center}
   \includegraphics[totalheight=1.9in,angle=0]{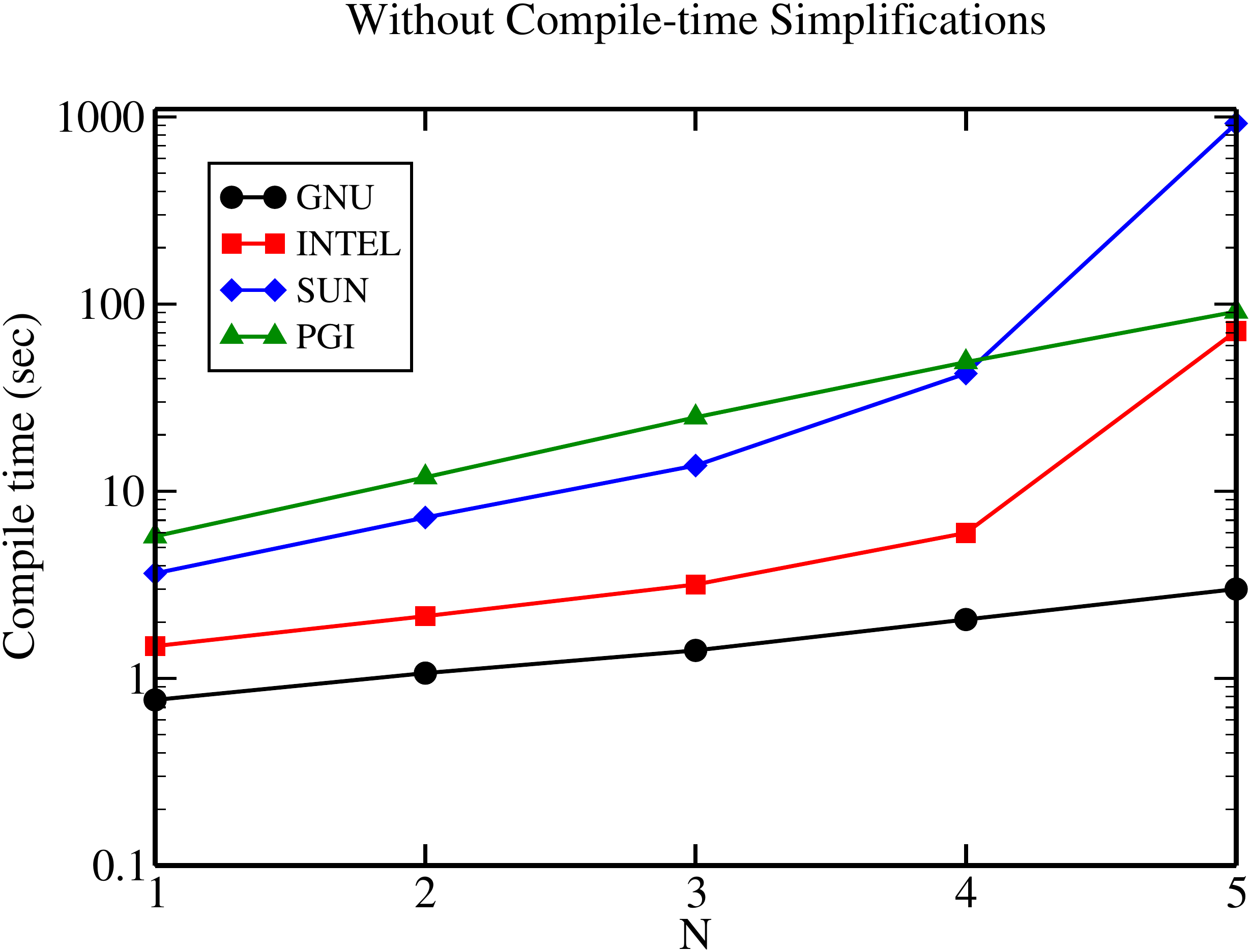}
   \includegraphics[totalheight=1.9in,angle=0]{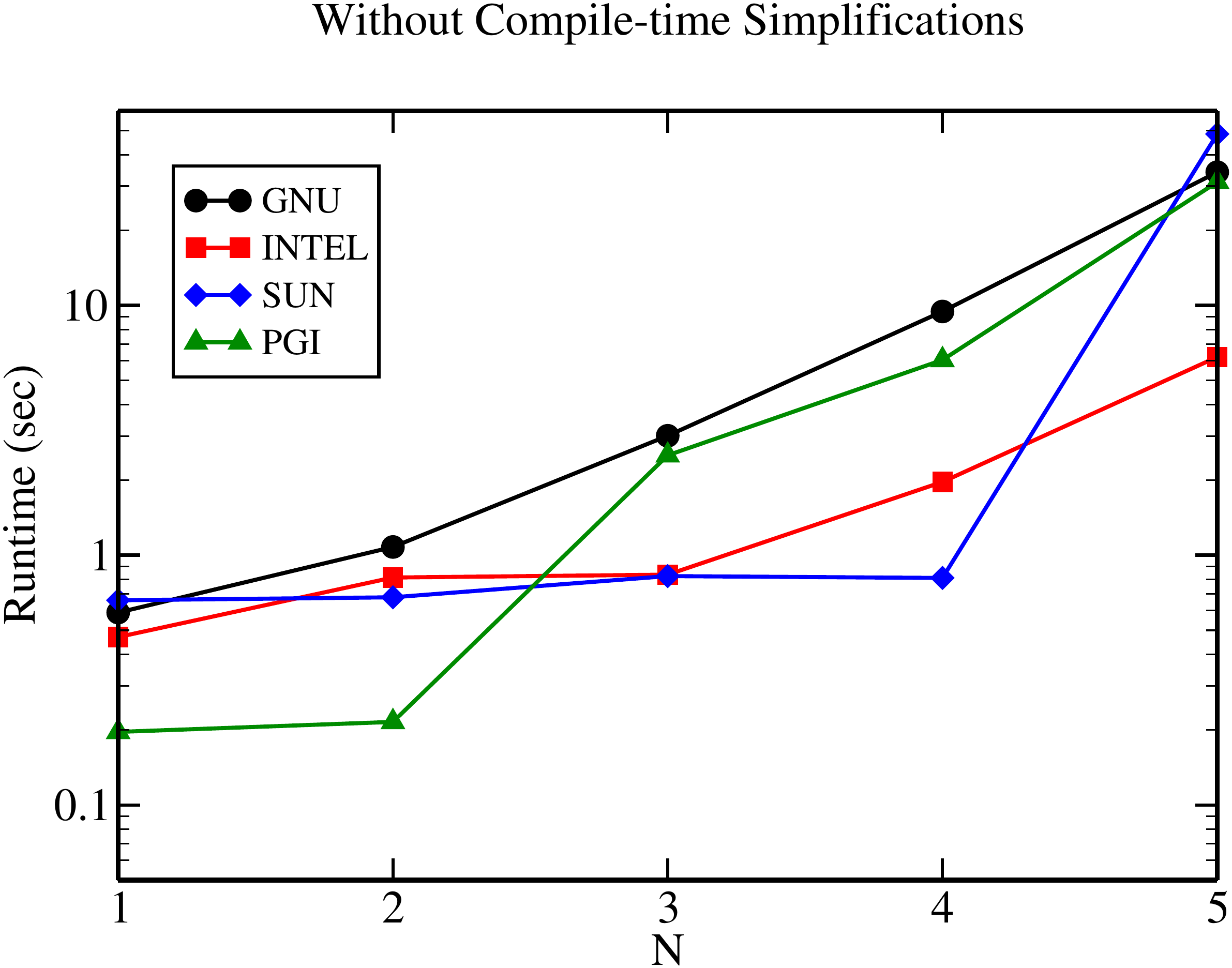}
\end{center} 
\caption{Compilation time needed by each compiler to generate the
  $N$th derivative of the function $f(x) = e^x + e^{2x} + e^{3x}$ when
  no simplifications are performed during compilation. 
  Runtime needed by each compiler-generated $N$th
   derivative of the function $f(x) = e^x + e^{2x} + e^{3x}$ for
    $10^7$ function calls.}
\label{fig:noSimplificationsInteger}
\end{figure}


At the left of Figure~\ref{fig:noSimplificationsInteger}, we plot the
compile time for each compiler, and at the right the runtime for
$10^7$ calls to the compiler-generated derivatives.  We see that all
compilers except \gnunew{}
 show rapidly increasing
compilation times and were not able to compile code for a derivative
of order five in reasonable time.  More precisely, the compilations
with \sun{} compiler required 922 seconds to compile the same code
where \gnunew{} required only 3 seconds. On the other had we observe
in Figure~\ref{fig:noSimplificationsInteger} right, that the corresponding runtimes
increase exponentially after a fourth-order  derivative.

\subsubsection{Simplifications enabled}
 
\begin{figure}[tb]   
  \begin{center}
    \includegraphics[totalheight=1.9in,angle=0]{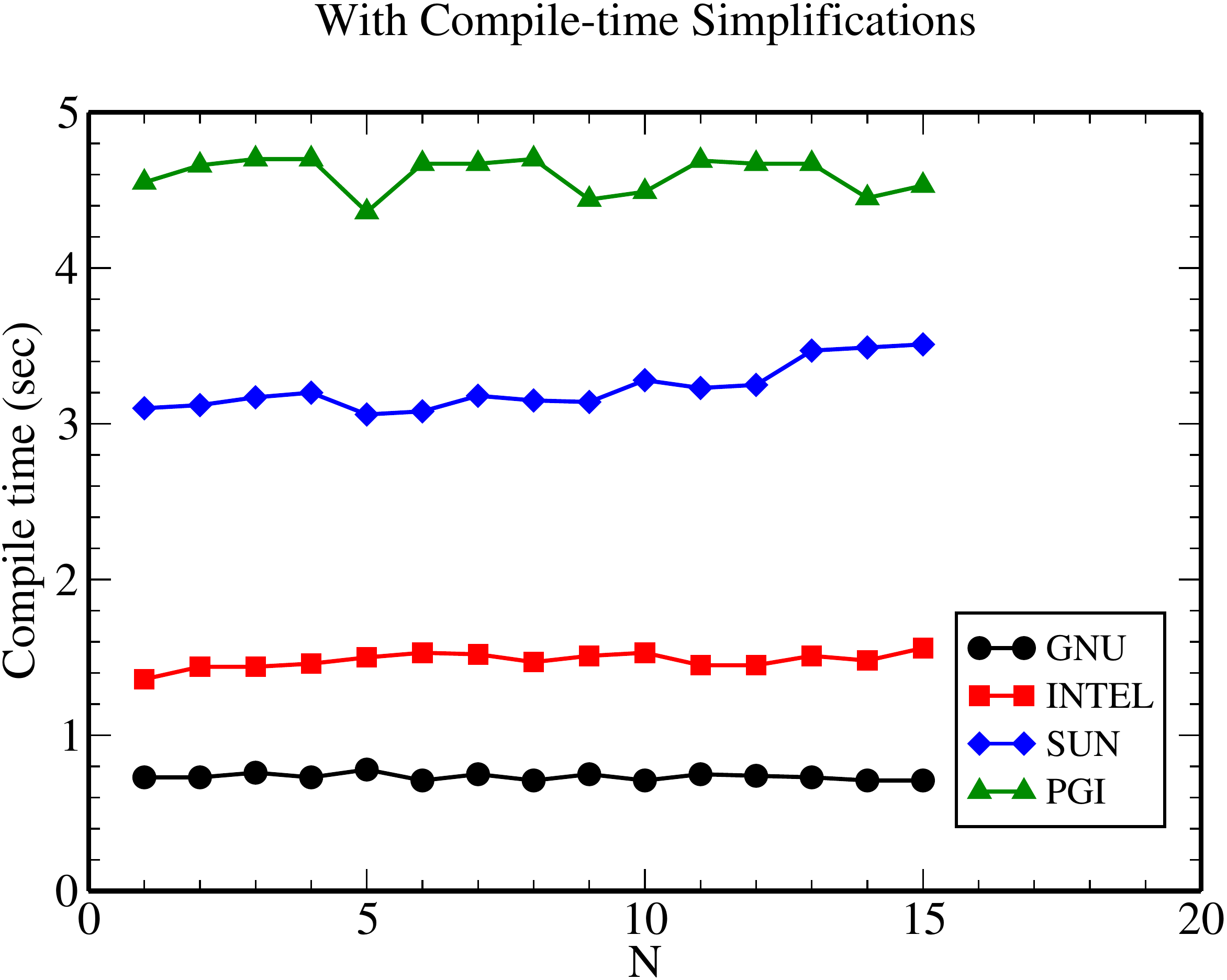}
    \includegraphics[totalheight=1.9in,angle=0]{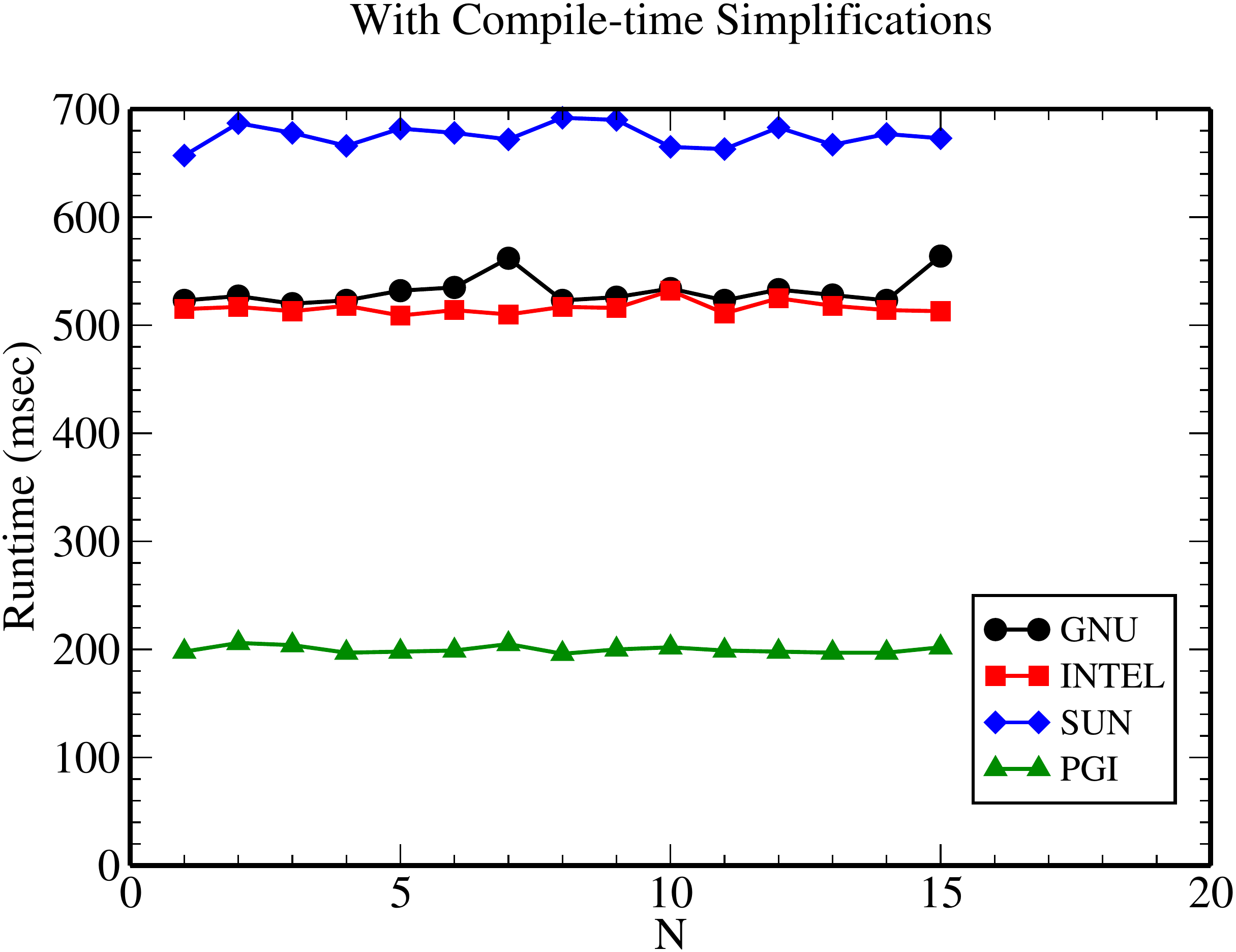}
  \end{center}
  \caption{Compilation time at the left needed by each compiler to generate
    $N$th derivatives of $f(x) = e^x + e^{2x} + e^{3x}$ when
        simplifications are performed during compilation. The run time for $10^7$ function calls
        of the compiler-generated $N$th
    derivative of the function $f(x)$ is shown at the right.}
  \label{fig:simplificationsInteger}
\end{figure}

We ran the same benchmark with simplifications enabled in our code.
The simplification code produces the same expression for the $N$th 
derivative as a hand coded version of derivative. More precicely the derivative expression
computed during compilation obtains the simplified form $d^Nf(x)/dx^N = e^x + 2^N e^{2x} + 3^N e^{3x}$.
The results are depicted in
Figure~\ref{fig:simplificationsInteger}. Compilation time and runtime of
the generated derivatives remain practically constant and independent of the
differentiation order. We were able to obtain derivatives up to order
15 before the integer constant coefficient ($3^N$) of some of the
exponential terms overflowed. The runtime here as well corresponds to $10^7$
function calls.

\subsection{Beyond template integer constants}
\label{ssec:template-constants}

As we discussed in \S\ref{sssec:constants-integer}, the \approachint{}
approach is not the optimal way to implement constants, as it can only
represent integer numbers.  It was used to illustrate the convenience
and benefits with respect to compile time simplifications that result
by its adoption.  The class \lstinline+Real<double Value>+, currently
not supported by the \Cpp{} language standard, would share the
flexibility of the class \lstinline+Integer<int Value>+ and would
provide the ideal way for implementing constants in our framework.
\begin{figure}[h]    
  \begin{center}
    \includegraphics[totalheight=1.8in,angle=0]{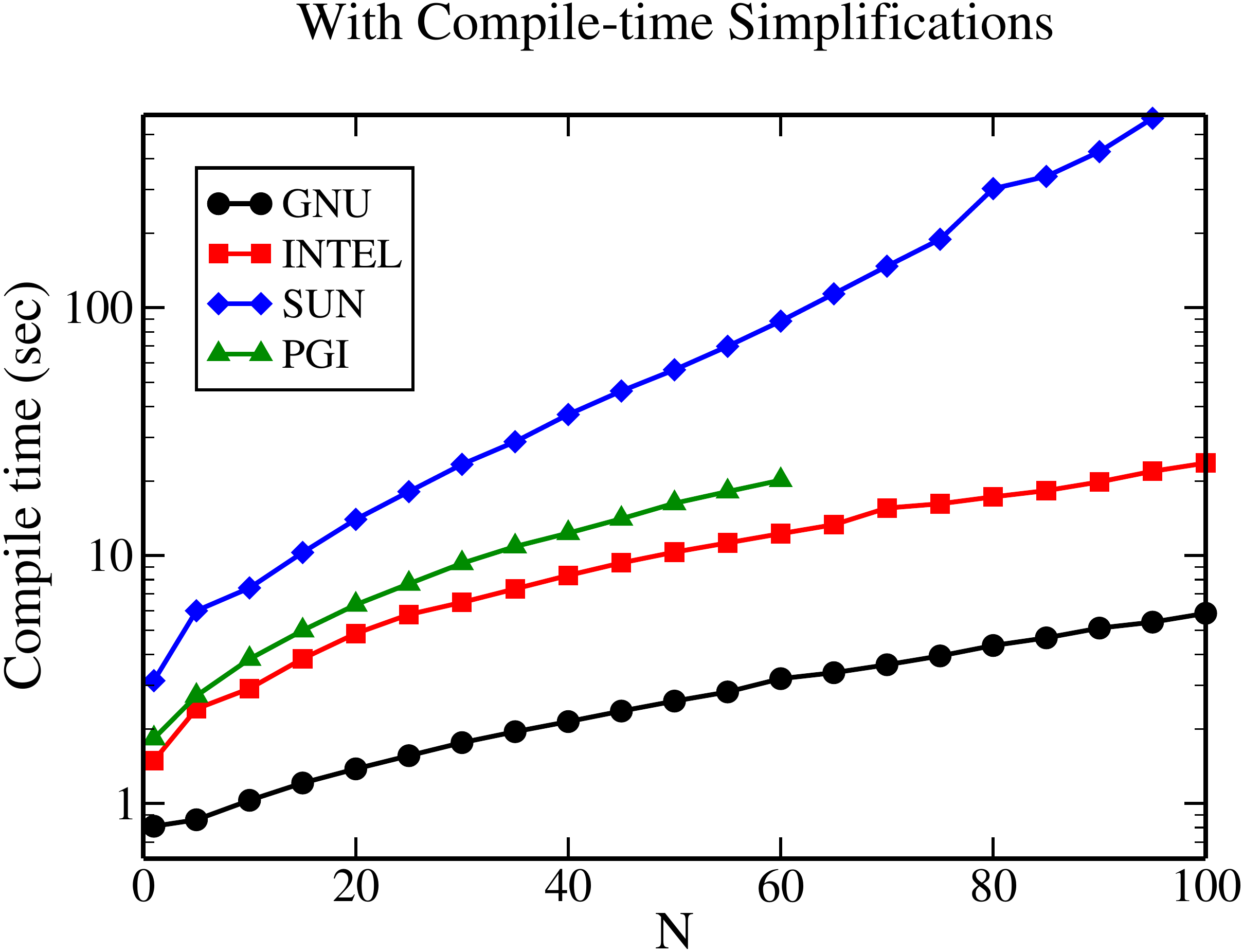}
    \includegraphics[totalheight=1.8in,angle=0]{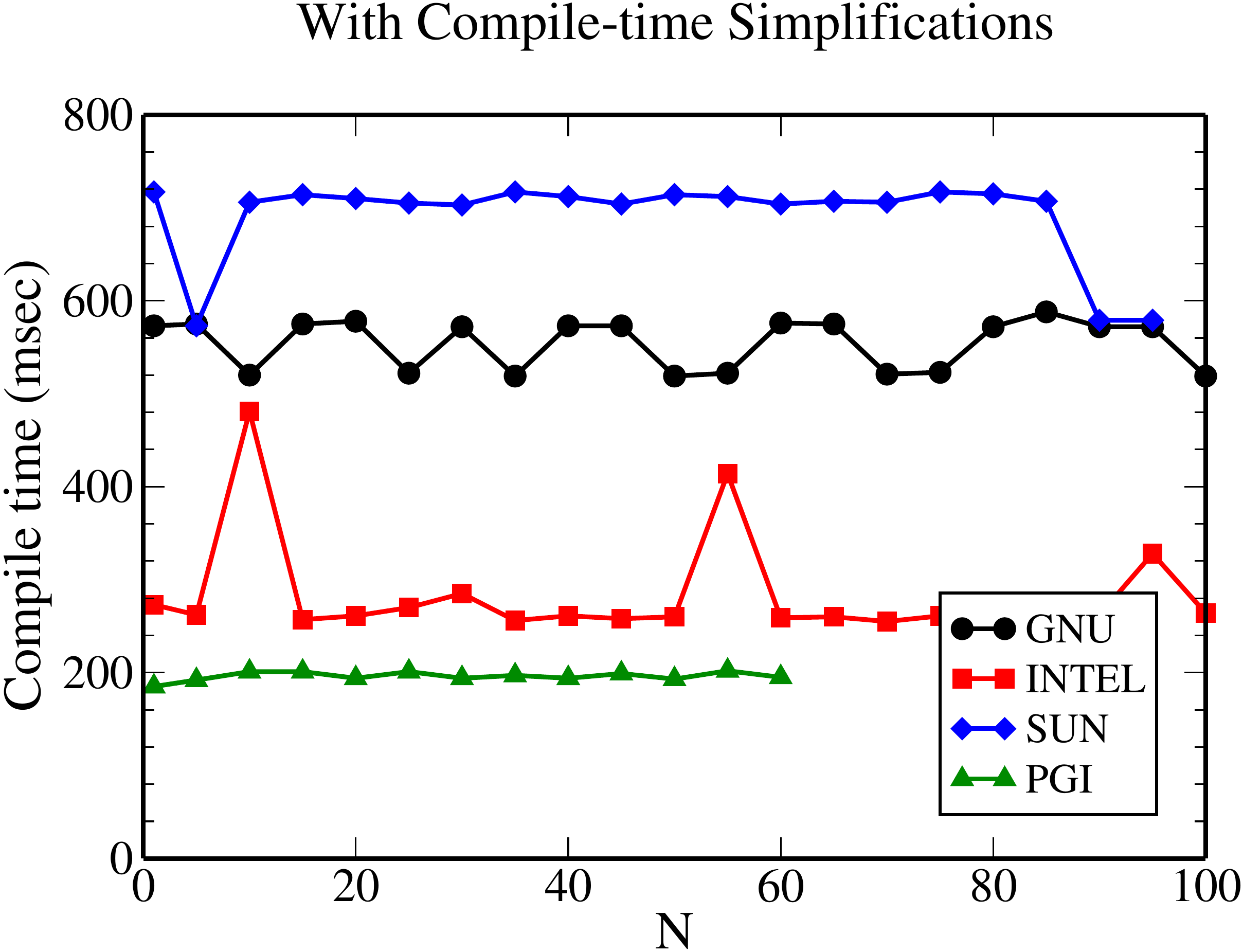}
  \end{center}
  \caption{Compilation time at the left, for obtaining the $N$th derivative of
    $f(x) = e^x + e^{2x} + e^{3x}$ when simplifications are performed
    during compilation with real constants implemented by the
    \approachreal{} approach.  Runtime for $10^7$ function calls 
    at the right, of the compiler-generated $N$th derivatives.}
\label{fig:simplificationsReal}
\end{figure}


Here we benchmark the only alternative way of implementing
constants, introduced in \S\ref{sssec:constants-real} and
\S\ref{sssec:constants-constant}.  Using the \approachreal{} approach
and with simplifications enabled we run the same benchmark as before,
but this time we can evaluate derivatives of much higher order than 15
because integer overflow is not an issue. Since the number of
objects that are generated with each new derivative increases
linearly, we expect the compilation time
to increase linearly as well.
Furthermore, the runtime should remain constant because
all the intermediate arithmetic operations between real constants are
wrapped by the \lstinline+class Constant<typename T>+ and calculated
upon construction of the object representing the derivative. Thus, no
redundant arithmetic operations are performed when the derivative is
evaluated. This is exactly what we observe in
Figure~\ref{fig:simplificationsReal}.


\subsection{Scalability for long formulae}

\begin{figure}[th]    
  \begin{center}
    \hbox to \textwidth{\hss
    \includegraphics[totalheight=1.9in,angle=0]{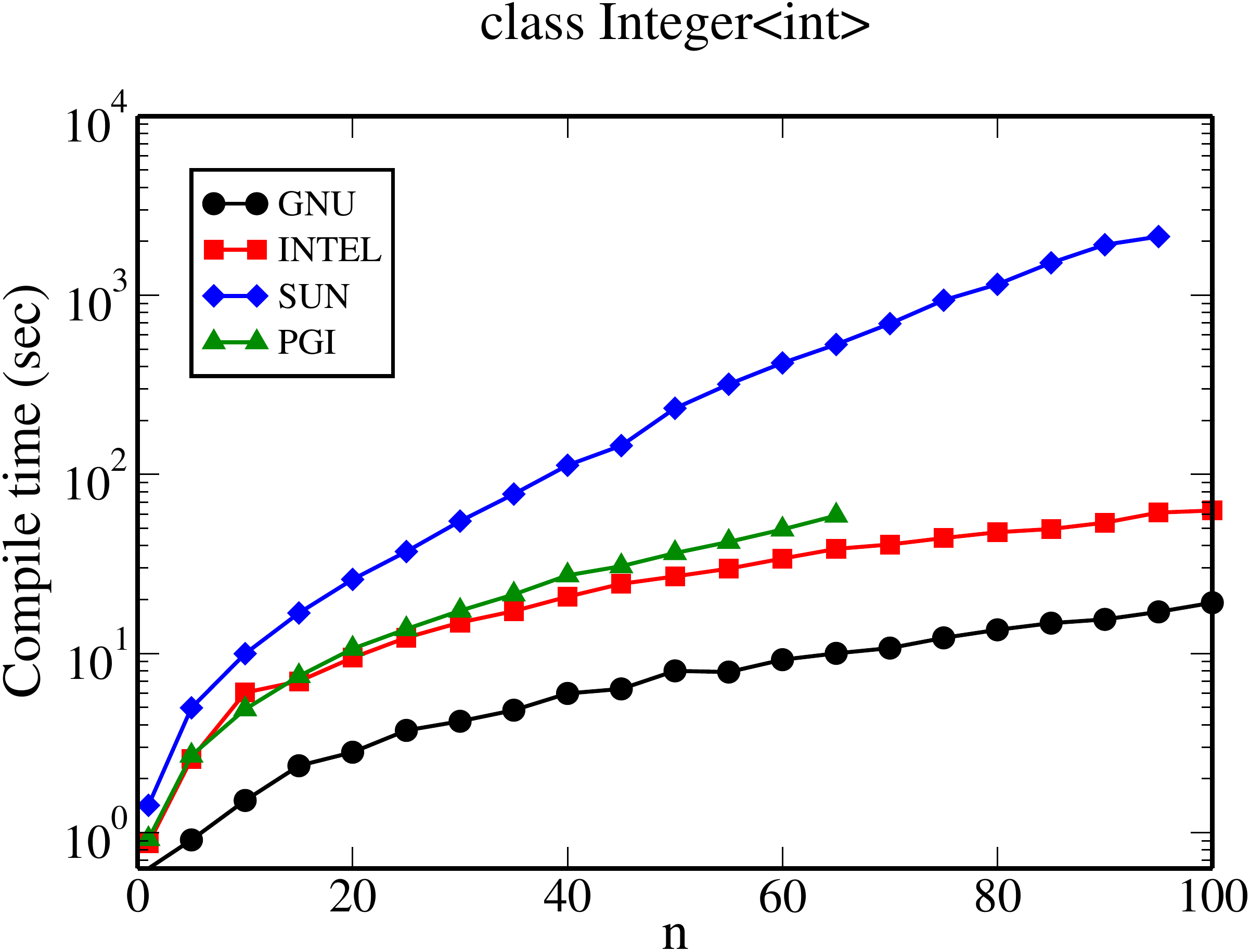}
    \includegraphics[totalheight=1.9in,angle=0]{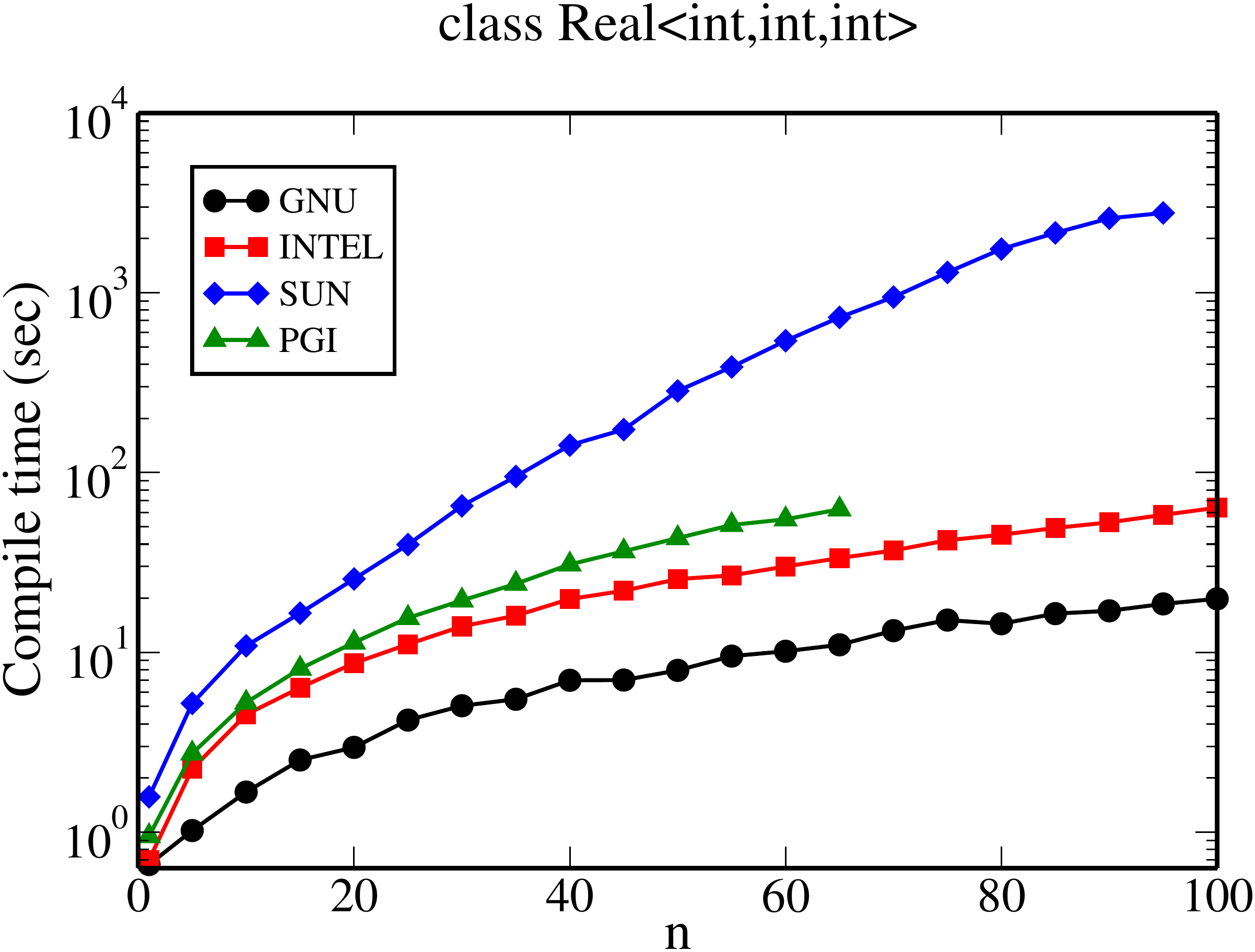}
                        \hss}
  \end{center}
  \caption{Compilation time for generating the 1st derivative of
    $f(x) = \sum_{j=1}^n e^{jx}$, with simplifications
    performed during compilation using the \approachint{} approach for
    constants (left) and the \approachreal{} approach
    (right).}
  \label{fig:scalabilityCompilationTime}
\end{figure}

Having established the benefits as well as the convenience and
flexibility that accompany compile time simplifications in both code
quality and runtime performance, we now study the
scalability of our approach for formulae consisting of many terms,
with simplifications turned on in our code.  For this purpose we
consider the following function of one variable:
\begin{align}
\label{eq:sumExp}
f(x) = \sum_{j=1}^{n} e^{jx},
\end{align}
for which our limited set of simplification rules works as intended.
By increasing the upper limit of the sum in (\ref{eq:sumExp}), we
obtain longer and longer expressions. As before, we examine both
approaches of implementing constants.  The compilation time
needed for obtaining the first derivative of (\ref{eq:sumExp}) with
both approaches is depicted in
Figure~\ref{fig:scalabilityCompilationTime}.

\begin{figure}[!tph]    
  \begin{center}
      \includegraphics[totalheight=1.9in,angle=0]{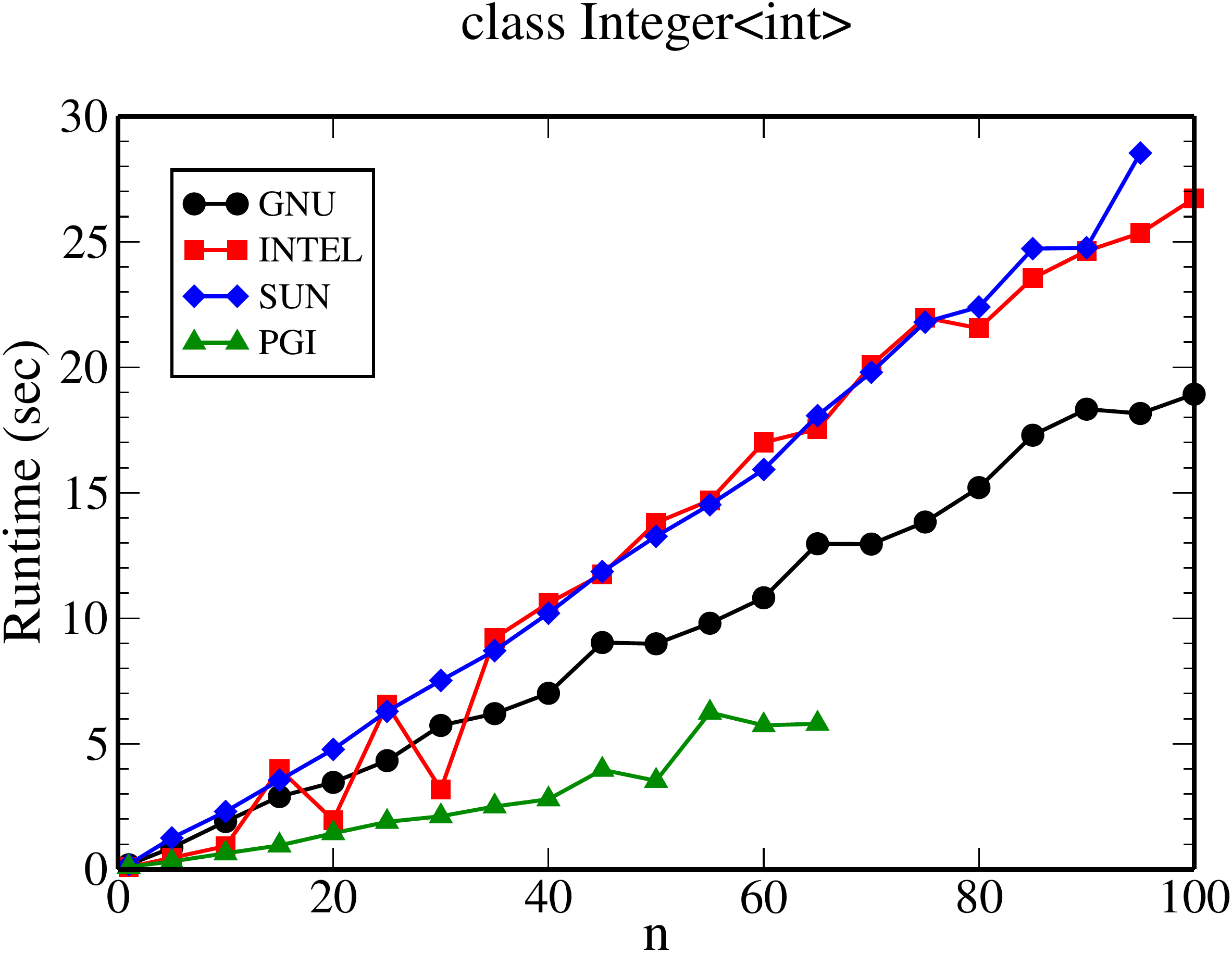}
    \includegraphics[totalheight=1.9in,angle=0]{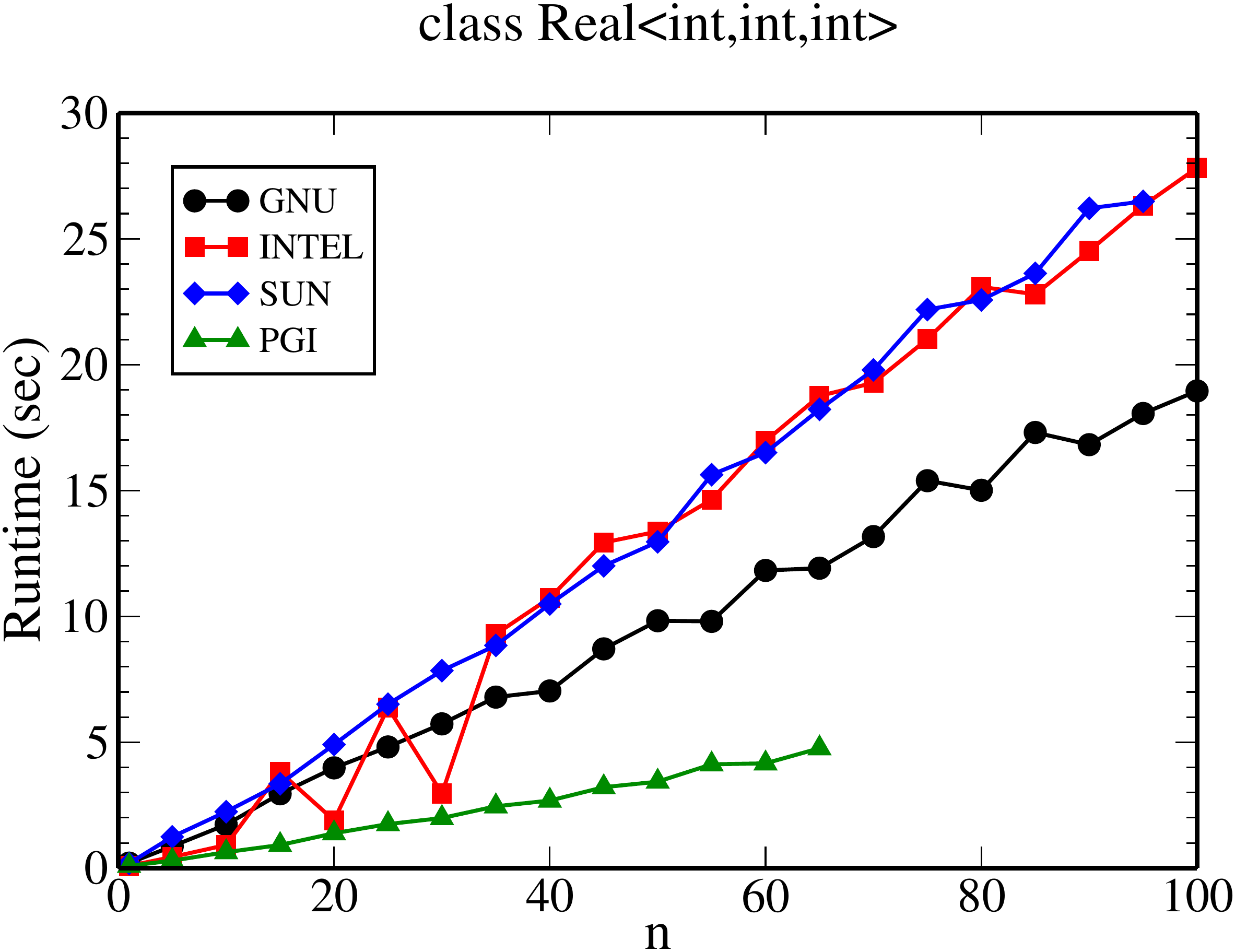}
  \end{center}
  \caption{Runtime of the generated 1st derivative of
    $f(x) = \sum_{j=1}^n e^{jx}$, with simplifications
    performed during compilation using the \approachint{} approach for
    constants (left) and the \approachreal{} approach
    (right), for $10^7$ function calls.}
  \label{fig:scalabilityRunTime} 
\end{figure}


The compilation time is depicted in
Figure~\ref{fig:scalabilityCompilationTime} with
constants implemented by the \approachint{} approach at the left
and with the real \approachreal{} at the right.  In
Figure~\ref{fig:scalabilityRunTime} we show the corresponding
runtimes of the compiler-generated derivatives.
We observe that the compilation time scales linearly for all compilers
with the exception of \sun{}, for which it increases exponentially.
Regarding the runtime performance, we see that for all compilers the
runtime increases also linearly. Moreoever it is the same as the hand-coded
one. 
This striking feature of our approach suggests that as soon as
simplification rules are mature enough to handle all possible cases,
then partial derivatives of any order obtained at compile-time would
perform as fast as the hand-coded ones. 


\section{Comparison with other AD approaches}    
\label{sec:otherAD}

It is of interest to compare the performance of the CoDET approach
against other popular AD packages like \FADBADpp{}
\cite{FadBadhtml,FadBadreport} and both the tape-based and tapeless
methods provided by ADOL-C \cite{Walther2012,Griewank,ADOLCBook}.  \FADBADpp{}
uses \Cpp{} expression templates while ADOL-C provides a library
to which the user should link after modifying appropriately his code.
The tapeless approach provided by ADOL-C is much more efficient than
the tape-based one, but it can only obtain first partial derivatives.  The \gnunew{} compiler was used throughout.

\subsection{Univariate functions}

\begin{figure}[th]    
  \begin{center}
    \includegraphics[totalheight=1.86in,angle=0]{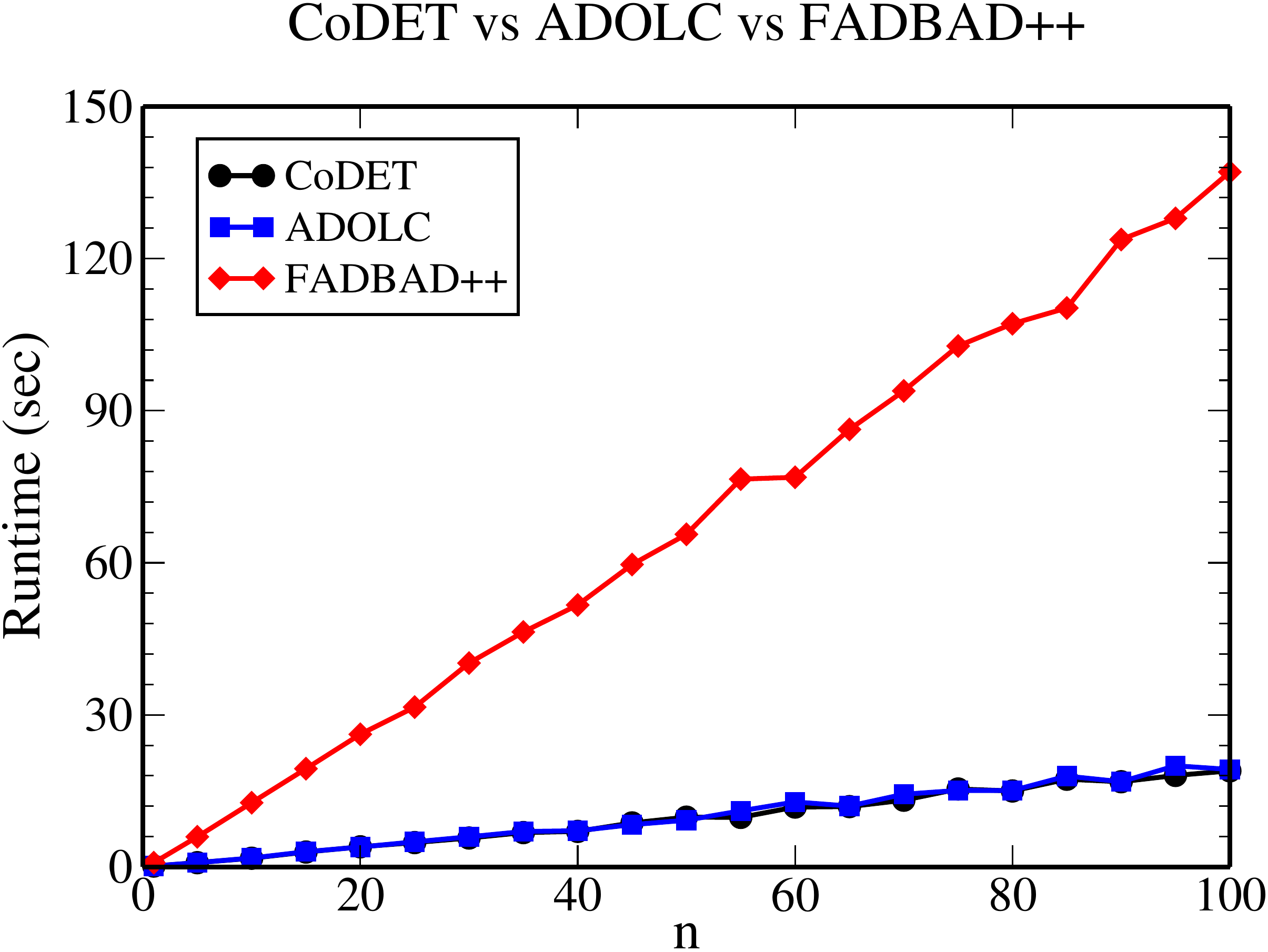}
    \includegraphics[totalheight=1.86in,angle=0]{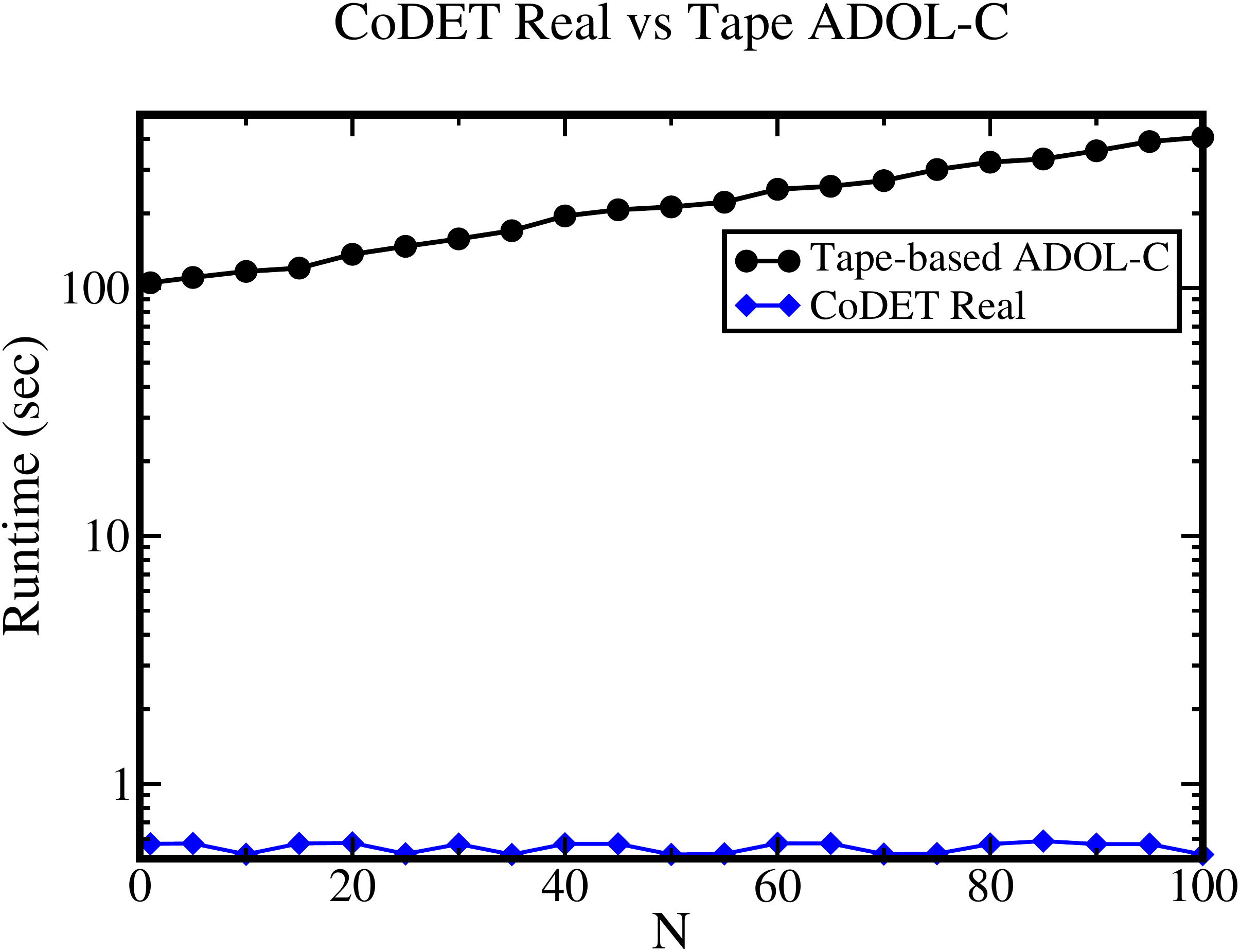}
  \end{center}
  \caption{Runtime of the compiler-generated first derivative of
    the function $f(x) = \sum_{j=1}^n e^{jx}$ obtained by our 
    CoDET \approachreal{} approach versus the one obtained by
      \protect\FADBADpp{}
      and both tape based and tapeless ADOL-C (left). 
      Runtime of the $N$th derivatives
      of $f(x) = e^x + e^{2x} + e^{3x}$ obtained by the CoDET \approachreal{} 
    approach versus tape-based ADOL-C (right). All runtimes correspond 
    to $10^7$ function calls.}
\label{fig:otherADSoftware}
\end{figure}

Our first benchmark considers the first derivative of
$f(x) = \sum_{j=1}^n e^{jx}$ in (\ref{eq:sumExp}) for $1 \le n \le 100$. 
The runtime for $10^7$ function calls as a function of
$n$ is plotted in Figure~\ref{fig:otherADSoftware} (left). We clearly see
that our approach and the tapeless ADOL-C approach are as fast as 
the hand-coded derivative, and even a bit faster for large $n$.
The remaining approaches are significantly slower.

At the right of Figure~\ref{fig:otherADSoftware} we compare the
runtime of the $N$th derivative of $f(x) = e^x + e^{2x} + e^{3x}$.
The CoDET runtime is constant and independent of the differentiation
order $N$, while the tape-based ADOL-C needs increasingly more time as
$N$ grows.  Thus our approach performs from 200 to 800 times
faster for $N=1$ to 100.
  
\subsection{Multivariate functions}

We now test the performance of CoDET with the two multivariate
functions
\begin{align}
   \label{eq:mvfg}
   f(\mathbf x) =  \frac{x_0 \tan(x_1 x_2)}{\tan(x_1 x_2) - x_3},
\qquad g(\mathbf x) =  x_0  + \sqrt{\sqrt{x_1 + \sqrt{x_2 + x_3}}}.
\end{align}
In both cases we evaluate all partial derivatives of $f(\mathbf x)$ and $g(\mathbf x)$.
The runtimes for $10^7$ function calls are listed in
Table~\ref{table:otherADSoftware}. Once again we see that our approach
is as efficient as derivatives coded by hand as ordinary
inline \Cpp{} functions. All other approaches lag behind in
performance. The fastest alternative (tapeless ADOL-C) is two times 
slower for the first case and demonstrates similar performance with
hand coded derivatives for all except the first partial derivative $\partial g / \partial x_0$
for which it performs 23 times slower than CoDET.

\begin{table} [h]   \newcommand{\z}{\phantom{0}}
\caption{Runtime of the generated partial derivatives
  of the multivariate functions $f$ and $g$ in equation \eqref{eq:mvfg}
  using several AD approaches:
  CoDET, \protect\FADBADpp, tape-based ADOL-C, and tapeless ADOL-C.
  The runtimes correspond to $10^7$ function calls.}
\hbox to \textwidth{\hss
\begin{tabular}{cccccc}
\hline
 & Hand-coded & CoDET & \FADBADpp & Tape ADOL-C & Tapeless ADOL-C \\ 
\midrule
  $\partial f/ \partial x_0$  &\z0.42 &\z0.42 & 3.54 & 154.85 & 0.78 \\
  $\partial f/ \partial x_1$  &\z0.45 &\z0.45 & 11.45 & 166.41&0.83 \\
  $\partial f/ \partial x_2$  &\z0.45 &\z0.45 & 11.64 & 159.73 & 0.8 \\
  $\partial f/ \partial x_3$  &\z0.41 &\z0.41 & 3.43 & 159.97 & 0.8 \\
\midrule
 $\partial g/ \partial x_0$  & 0.01 & 0.01 & 1.12 & 156.62 & 0.23 \\
 $\partial g/ \partial x_1$  & 0.51 & 0.39 & 5.52 & 156.51 & 0.58 \\
 $\partial g/ \partial x_2$  & 0.59 & 0.46 & 3.97 & 159.71& 0.61 \\
 $\partial g/ \partial x_3$  & 0.59 & 0.46 & 3.97 & 159.08& 0.61\\
 \midrule
\end{tabular}\hss}
\label{table:otherADSoftware}
\end{table}
We understand that the performance depends on the specific function
being benchmarked.  However, provided our library is augmented with
sophisticated simplification compile time rules to handle every
possible case (which is not our purpose here), our approach should
always produce derivatives as efficient as hand-coded derivatives.
The \Cpp{} code needed for encoding $f(\mathbf x)$ as a template
expression and generating its first derivative using the \auto{}
keyword follows:
%
%
%
%
%
%
\begin{lstlisting}
Variable<0> X0; Variable<1> X1;
Variable<2> X2; Variable<3> X3;

auto f = ( X0*tan(X1*X2) )/( tan(X1*X2)-X3 );

const int n = 1;
const int m = 0;
auto dnf_dx_mn = f<n, m>.derivative();
\end{lstlisting}

Our final benchmark repeats the first computational experiment presented
by Nehmeier~\cite{Nehmeier}. Since the results presented in~\cite{Nehmeier} have
been obtained on a different system with an older compiler, 
we normalise the computation of gradients obtained from several different 
libraries with the runtimes of the hand-coded gradient. We provide only
the actual running time for the hand-coded gradient in milliseconds corresponding
to $10^7$ function calls. The results are presented in table~\ref{table:Nehmeier}.
The final row at the table shows the ratio of the running times of the compiler 
generated gradient using the approach introduced by Nehmeier~\cite{Nehmeier} with the 
hand coded versions that are reported there. As expected the CoDET approach has identical
running times with optimised hand-coded derivatives, unlike other competitors. 
Although the approach introduced in~\cite{Nehmeier} is similar to CoDET, it lacks
the simplification mechanism that is exploited by CoDET to avoid redundant computations.
\begin{table} [ht]
\centering
\caption{Performance comparison of the gradient computation. Numbers are normalised with the 
runtime of hand-coded gradients, measured in milliseconds, corresponding to $10^7$ functions calls.}
\begin{tabular*}{150mm}{@{\extracolsep\fill}lccc}\hline
Run                & $x^2y^3 + y\log(x)$         & $3x^2y-y^3$  & $(1-x)^2 + 100(y-x^2)$\\[2pt]
\midrule
Hand-coded (ms)    & 118           &  27         &       27   \\
\hline
CoDET              & 1             &  1          &       1      \\
FADBAD++           & 65.7          &  231.9      &       284.4  \\
ADOL-C             & 311           &  1300       &      1466.7  \\
ADOL-C reuse tape & 168.6         &  692.6      &       744.4  \\
Sacado DFad        & 8.3           &  27.7       &       39.2   \\
Sacado SFad        & 3             &  1.8        &       2.4    \\
Nehmeier           & 1.1     &  1.3        &       1.8    \\ 
\midrule
\end{tabular*}
  \label{table:Nehmeier}
\end{table}

\section{Conclusions}     

In our development of CoDET we have demonstrated how \Cpp{} expression
templates and template metaprogramming techniques can be employed to
allow \Cpp{} compilers to generate partial derivatives of multivariate
functions of any order during the compilation process. We verified
that compile time simplifications of the resulting formulas for the
derivatives must be interleaved with the differentiation steps in
order to speed up compilation.  For some cases, the implementation of
compile time simplification rules resulted in a reduction of
compilation time by up to three orders of magnitude for specific
compilers, while the runtime of the generated derivatives was reduced
by up to two orders of magnitude.

A striking feature of our approach, apart from the
arbitrarily high order of derivatives that can be obtained, is that
the compiler-generated derivatives are as efficient as hand-coded
ones, provided a complete set of simplification rules is implemented.

The template metaprogramming techniques presented and benchmarked
here revealed that several \Cpp{} compilers are already mature
enough for compile time symbolic differentiation.  The same techniques
could also be used to implement symbolic integration at compile time.
We hope that this work will motivate further developments by compiler
vendors with the aim of supporting complete symbolic compile time
differentiation in \Cpp{} and other high-level languages.




\bibliographystyle{plain}
\bibliography{CompileTimeSymbolicAD.bib}

\end{document}

%% file: est1.pstex_t
\begin{picture}(0,0)%
\includegraphics{est1.pdf}%
\end{picture}%
\setlength{\unitlength}{3947sp}%
\begingroup\makeatletter\ifx\SetFigFont\undefined%
\gdef\SetFigFont#1#2#3#4#5{%
  \reset@font\fontsize{#1}{#2pt}%
  \fontfamily{#3}\fontseries{#4}\fontshape{#5}%
  \selectfont}%
\fi\endgroup%
\begin{picture}(2487,1925)(2925,-4161)
\put(3064,-4027){\makebox(0,0)[lb]{\smash{{\SetFigFont{12}{14.4}{\familydefault}{\mddefault}{\updefault}{\color[rgb]{0,0,0}$2$}%
}}}}
\put(3301,-3506){\makebox(0,0)[lb]{\smash{{\SetFigFont{12}{14.4}{\familydefault}{\mddefault}{\updefault}{\color[rgb]{0,0,0}$\times $}%
}}}}
\put(3539,-4044){\makebox(0,0)[lb]{\smash{{\SetFigFont{12}{14.4}{\familydefault}{\mddefault}{\updefault}{\color[rgb]{0,0,0}$x_2$}%
}}}}
\put(4351,-4041){\makebox(0,0)[lb]{\smash{{\SetFigFont{12}{14.4}{\familydefault}{\mddefault}{\updefault}{\color[rgb]{0,0,0}$x_0$}%
}}}}
\put(4951,-4035){\makebox(0,0)[lb]{\smash{{\SetFigFont{12}{14.4}{\familydefault}{\mddefault}{\updefault}{\color[rgb]{0,0,0}$x_1$}%
}}}}
\put(4651,-3506){\makebox(0,0)[lb]{\smash{{\SetFigFont{12}{14.4}{\familydefault}{\mddefault}{\updefault}{\color[rgb]{0,0,0}$\times $}%
}}}}
\put(4576,-2919){\makebox(0,0)[lb]{\smash{{\SetFigFont{12}{14.4}{\familydefault}{\mddefault}{\updefault}{\color[rgb]{0,0,0}exp}%
}}}}
\put(3986,-2462){\makebox(0,0)[lb]{\smash{{\SetFigFont{12}{14.4}{\familydefault}{\mddefault}{\updefault}{\color[rgb]{0,0,0}$+$}%
}}}}
\end{picture}%